\documentclass[acmsmall,screen]{acmart}

\renewcommand\footnotetextcopyrightpermission[1]{}

\usepackage{booktabs}   
\usepackage{makecell}   
\usepackage{colortbl}   
\usepackage{tikz}       
\usepackage{pgfplots}   
\pgfplotsset{compat=1.18} 
\usepackage{pgf-pie}
\usepackage{subcaption}
\usepackage{float}

\usepackage{tabularx}
\usepackage{placeins}
\usepackage{multirow}
\usepackage{graphicx}

\usepackage{array}
\usepackage{ragged2e}

\usepackage{caption}

\AtBeginDocument{%
  }

\newcolumntype{P}[1]{>{\raggedright\arraybackslash}p{#1}}

\expandafter\def\csname redditmap@1qnntl7\endcsname{R001}
\expandafter\def\csname redditmap@1qo21fg\endcsname{R002}
\expandafter\def\csname redditmap@1qo8hmo\endcsname{R003}
\expandafter\def\csname redditmap@1qoqsah\endcsname{R004}
\expandafter\def\csname redditmap@1qoxsjj\endcsname{R005}
\expandafter\def\csname redditmap@1qr8njo\endcsname{R006}
\expandafter\def\csname redditmap@1qrkhdo\endcsname{R007}
\expandafter\def\csname redditmap@1qs0z11\endcsname{R008}
\expandafter\def\csname redditmap@1qsg7mj\endcsname{R009}
\expandafter\def\csname redditmap@1qstmrl\endcsname{R010}
\expandafter\def\csname redditmap@1qsy4gi\endcsname{R011}
\expandafter\def\csname redditmap@1qta1fb\endcsname{R012}
\expandafter\def\csname redditmap@1qtb7b0\endcsname{R013}
\expandafter\def\csname redditmap@1qtmk1c\endcsname{R014}
\expandafter\def\csname redditmap@1qto1yd\endcsname{R015}
\expandafter\def\csname redditmap@1qtux4c\endcsname{R016}
\expandafter\def\csname redditmap@1qu4wy8\endcsname{R017}
\expandafter\def\csname redditmap@1qu5rw8\endcsname{R018}
\expandafter\def\csname redditmap@1qu8h73\endcsname{R019}
\expandafter\def\csname redditmap@1quicq4\endcsname{R020}
\expandafter\def\csname redditmap@1qujfnn\endcsname{R021}
\expandafter\def\csname redditmap@1qum66s\endcsname{R022}
\expandafter\def\csname redditmap@1qvjxkn\endcsname{R023}
\expandafter\def\csname redditmap@1qvogps\endcsname{R024}
\expandafter\def\csname redditmap@1qwoi9q\endcsname{R025}
\expandafter\def\csname redditmap@1qww2ih\endcsname{R026}
\expandafter\def\csname redditmap@1qwx0z1\endcsname{R027}
\expandafter\def\csname redditmap@1qx1rqd\endcsname{R028}
\expandafter\def\csname redditmap@1qx58u2\endcsname{R029}
\expandafter\def\csname redditmap@1qxapke\endcsname{R030}
\expandafter\def\csname redditmap@1qxq3la\endcsname{R031}
\expandafter\def\csname redditmap@1qy56j9\endcsname{R032}
\expandafter\def\csname redditmap@1qy8v32\endcsname{R033}
\expandafter\def\csname redditmap@1qyado7\endcsname{R034}
\expandafter\def\csname redditmap@1qycx0r\endcsname{R035}
\expandafter\def\csname redditmap@1qywzks\endcsname{R036}
\expandafter\def\csname redditmap@1qz6wzs\endcsname{R037}
\expandafter\def\csname redditmap@1qzl3u2\endcsname{R038}
\expandafter\def\csname redditmap@1qzx49g\endcsname{R039}
\expandafter\def\csname redditmap@1r06z0t\endcsname{R040}
\expandafter\def\csname redditmap@1r29h8t\endcsname{R041}
\expandafter\def\csname redditmap@1r2dygf\endcsname{R042}
\expandafter\def\csname redditmap@1r2enjm\endcsname{R043}
\expandafter\def\csname redditmap@1r2jbtz\endcsname{R044}
\expandafter\def\csname redditmap@1r2lfrl\endcsname{R045}
\expandafter\def\csname redditmap@1r2xxtn\endcsname{R046}
\expandafter\def\csname redditmap@1r33nkn\endcsname{R047}
\expandafter\def\csname redditmap@1r39q77\endcsname{R048}
\expandafter\def\csname redditmap@1r51mb6\endcsname{R049}
\expandafter\def\csname redditmap@1r5403m\endcsname{R050}
\expandafter\def\csname redditmap@1r6g3w6\endcsname{R051}
\expandafter\def\csname redditmap@1r6nwyz\endcsname{R052}
\expandafter\def\csname redditmap@1r6sk3a\endcsname{R053}
\expandafter\def\csname redditmap@1r6uy26\endcsname{R054}
\expandafter\def\csname redditmap@1r6y0g6\endcsname{R055}
\expandafter\def\csname redditmap@1r6z47f\endcsname{R056}
\expandafter\def\csname redditmap@1r76n9u\endcsname{R057}
\expandafter\def\csname redditmap@1r77ynv\endcsname{R058}
\expandafter\def\csname redditmap@1r77zso\endcsname{R059}
\expandafter\def\csname redditmap@1r7boaz\endcsname{R060}
\expandafter\def\csname redditmap@1r7r3nm\endcsname{R061}
\expandafter\def\csname redditmap@1r85hj5\endcsname{R062}
\expandafter\def\csname redditmap@1r86v5e\endcsname{R063}
\expandafter\def\csname redditmap@1r8nnnc\endcsname{R064}
\expandafter\def\csname redditmap@1r8os2i\endcsname{R065}
\expandafter\def\csname redditmap@1r97qia\endcsname{R066}
\expandafter\def\csname redditmap@1r99j4h\endcsname{R067}
\expandafter\def\csname redditmap@1r9lx3a\endcsname{R068}
\expandafter\def\csname redditmap@1r9pwov\endcsname{R069}
\expandafter\def\csname redditmap@1r9tqj9\endcsname{R070}
\expandafter\def\csname redditmap@1ra1hzn\endcsname{R071}
\expandafter\def\csname redditmap@1radcrw\endcsname{R072}
\expandafter\def\csname redditmap@1ratn8g\endcsname{R073}
\expandafter\def\csname redditmap@1rawxuz\endcsname{R074}
\expandafter\def\csname redditmap@1rboqul\endcsname{R075}
\expandafter\def\csname redditmap@1rc1mei\endcsname{R076}
\expandafter\def\csname redditmap@1rcisy3\endcsname{R077}
\expandafter\def\csname redditmap@1rco0fp\endcsname{R078}
\expandafter\def\csname redditmap@1rd7zyq\endcsname{R079}
\expandafter\def\csname redditmap@1rdeb75\endcsname{R080}
\expandafter\def\csname redditmap@1rdolnf\endcsname{R081}
\expandafter\def\csname redditmap@1reetho\endcsname{R082}
\expandafter\def\csname redditmap@1rejuyn\endcsname{R083}
\expandafter\def\csname redditmap@1rekom1\endcsname{R084}
\expandafter\def\csname redditmap@1rer6l3\endcsname{R085}
\expandafter\def\csname redditmap@1resh66\endcsname{R086}
\expandafter\def\csname redditmap@1revd0q\endcsname{R087}
\expandafter\def\csname redditmap@1rewrct\endcsname{R088}
\expandafter\def\csname redditmap@1rg3s8j\endcsname{R089}
\expandafter\def\csname redditmap@1rg8qbc\endcsname{R090}
\expandafter\def\csname redditmap@1rgcunb\endcsname{R091}
\expandafter\def\csname redditmap@1rgklwl\endcsname{R092}
\expandafter\def\csname redditmap@1rhrwlz\endcsname{R093}
\expandafter\def\csname redditmap@1ri3582\endcsname{R094}
\expandafter\def\csname redditmap@1ri8bgc\endcsname{R095}

\newcommand{\rlnk}[1]{%
  \href{https://www.reddit.com/comments/#1/}{%
    \ifcsname redditmap@#1\endcsname
      \csname redditmap@#1\endcsname
    \else
      R-UNK
    \fi
  }%
}

\newcommand{\glnk}[1]{\href{https://github.com/openclaw/openclaw/issues/#1/}{\##1}}

\newcounter{observation}

\newcommand{\observation}[1]{%
  \refstepcounter{observation}%
  \par\noindent
  {\setlength{\fboxsep}{2pt}%
   \setlength{\fboxrule}{0.4pt}%
   \fbox{%
     \parbox{\dimexpr\linewidth-2\fboxsep-2\fboxrule\relax}{%
       \colorbox{black!80}{\textcolor{white}{\textbf{O\theobservation}}}%
       \hspace{0.5em}#1%
     }%
   }%
  }%
  \par\vspace{-0.4em}%
}

\newcounter{rec}

\newcounter{mitigation}
\newcommand{\mitfinding}[1]{%
  \refstepcounter{mitigation}%
  \begin{mdframed}[
    linewidth=0.4pt,
    roundcorner=2pt,
    innertopmargin=1pt,
    innerbottommargin=1pt,
    innerleftmargin=1pt,
    innerrightmargin=1pt,
    skipabove=1pt,
    skipbelow=1pt
  ]
  \noindent
  \colorbox{black!80}{%
    \textcolor{white}{\textbf{M\themitigation}}%
  }\hspace{0.5em}%
  #1
  \end{mdframed}
}
\usepackage{soul}
\usepackage[table]{xcolor}
\usepackage{xcolor}      
\usepackage{tcolorbox}   
\usepackage{mdframed}    
\usepackage{xparse}      
\usepackage{cuted}
\usepackage{caption}

\definecolor{quotegray}{RGB}{248,248,248}

\NewDocumentCommand{\evidencequote}{m m}{%
  \vspace{0.15em}
  \begin{mdframed}[
    backgroundcolor=quotegray,
    linecolor=quotegray,
    linewidth=0pt,
    leftline=false,
    rightline=false,
    topline=false,
    bottomline=false,
    innerleftmargin=2pt,
    innerrightmargin=2pt,
    innertopmargin=1pt,
    innerbottommargin=1pt,
    skipabove=0pt,
    skipbelow=0pt
  ]
  \noindent\small\raggedright\sloppy\emph{``#1''}~--~#2
  \end{mdframed}
}

\newtcolorbox{takeaway}[1][]{
    colback=blue!10!white,
    colframe=black,
    rounded corners,
    arc=3pt,
    boxrule=0.8pt,
    left=4pt,
    right=4pt,
    top=1pt,
    bottom=1pt,
    before skip=2pt,
    after skip=2pt,
    #1
}

\newcounter{rtext}

\usepackage{balance}
\newcommand{\commentout}[1]{}

\makeatletter
\newcommand{\reddittext}[2]{%
  ``\textit{\textcolor{black!85}{#1}}''--#2 
  \@afterindentfalse\@afterheading
}

\definecolor{lightshade}{RGB}{245,245,245} 



\begin{document}

\title{\textit{``Then there's prompt injection, a malicious email, ...''}: The Wild Wide West of OpenClaw Agentic System Security}


\author{Mostafijur Rahman Akhond}
\affiliation{%
  \institution{York University}
  \city{Toronto}
  \country{Canada}}
\email{mostafij@yorku.ca}

\author{Mahmuda Khatun}
\affiliation{%
  \institution{York University}
  \city{Toronto}
  \country{Canada}}
\email{mahmuda@yorku.ca}

\author{Gias Uddin}
\affiliation{%
  \institution{York University}
  \city{Toronto}
  \country{Canada}}
\email{guddin@yorku.ca}



\begin{abstract}
AI-powered personal assistant frameworks such as OpenClaw combine LLM reasoning, persistent memory, tool invocation, local execution, and web-based workflows. These capabilities enable useful automation, but they also create security risks when untrusted inputs, model reasoning, tools, memory, and deployment configuration interact without explicit trust boundaries. This paper presents an empirical study of reported security concerns in OpenClaw using two complementary data sources: Reddit discussions and security-tagged GitHub issues. We analyze more than 25K Reddit posts from 13 OpenClaw-related subreddits and identify 162 security-relevant posts with 1,787 associated comments. We further collect 358 security-tagged GitHub issues and retain 180 issues describing concrete security concerns. Using qualitative analysis and evidence-grounded CWE mapping, we identify 521 CWE instances across 59 distinct weakness types, concentrated in access-control failures, sensitive-data exposure, injection, and unsafe execution. We also extract 543 practitioner mitigation suggestions and consolidate them into seven recurring strategies. To complement the empirical analysis, we present an evidence-grounded illustrative reconstruction of a representative sensitive-data exposure pathway using dummy secret values. Our findings show that OpenClaw security issues are not isolated implementation bugs; they reflect recurring trust-boundary failures across the agent execution pipeline. We discuss how these weaknesses map to both traditional software-security categories and agent-specific failure mechanisms, and derive implications for safer agentic systems, including explicit trust boundaries, least-privilege tool access, execution containment, runtime oversight, and defense-in-depth deployment.
\end{abstract}




\maketitle

\section{Introduction}

Large Language Models (LLMs) have rapidly expanded beyond standalone chat-based assistants into autonomous, system-integrated agents that support complex workflows~\cite{dong2023towards, ha2024towards}. These systems combine LLM reasoning with external tools, persistent memory, and system-level capabilities, enabling multi-step task execution and continuous operation~\cite{li2024personal, li2025hello}. OpenClaw~\cite{openclaw} exemplifies this paradigm as a self-hosted, tool-integrated assistant with persistent state, tool access, and direct interaction with local and web-based resources. Unlike traditional LLM applications that generate isolated responses, such agentic systems can interact with files, execute commands, invoke external services, and retain context across sessions. These capabilities enable useful automation, but they also introduce security risks when untrusted inputs, model reasoning, tools, memory, and deployment configuration interact without explicit trust boundaries~\cite{chen2026trajectory, ying2026uncovering, shan2026don}.

Many resulting failures resemble familiar software weaknesses, including improper access control, data exposure, unsafe execution, and input validation errors. Prior work identifies key risks in LLM-based and agentic systems, such as prompt injection, unsafe tool use, privilege escalation, malicious skill execution, memory manipulation, and cross-context data exposure~\cite{deng2025ai, kim2026attack, liu2025make}. However, much of this work examines risks in controlled settings, through predefined attack scenarios, or at the level of isolated components such as model outputs, generated code, or benchmark tasks~\cite{chen2026lps, kim2025llms, ayzenshteyn2025cloak, he2025security}. In deployed agent frameworks, failures often emerge from composition: untrusted inputs influence model reasoning, model outputs trigger tools, tools access local or external resources, and persistent state carries information across interactions. Thus, agentic-system security cannot be fully understood by studying the model or tool layer alone.

OpenClaw provides a timely case for studying these risks in practice. Public vulnerability records and community reports suggest that such failures are already materializing: \textbf{CVE-2026-24763} reports an authenticated command-injection weakness in OpenClaw/Clawdbot Docker execution~\cite{cve2026openclaw}, while public reports document related agent-platform failures involving exposed private messages, credentials, unintended destructive actions, unauthorized data access, and misconfigured deployments~\cite{cna2026moltbook, toi2026openclaw_inbox, guardian2026openclaw, wired2026clawdbot}. These reports motivate an empirical view of OpenClaw security that asks how security concerns are reported, categorized, and mitigated across a real-world agent ecosystem.

\begin{figure}[t]
\centering
\includegraphics[height=0.35\textheight,keepaspectratio]{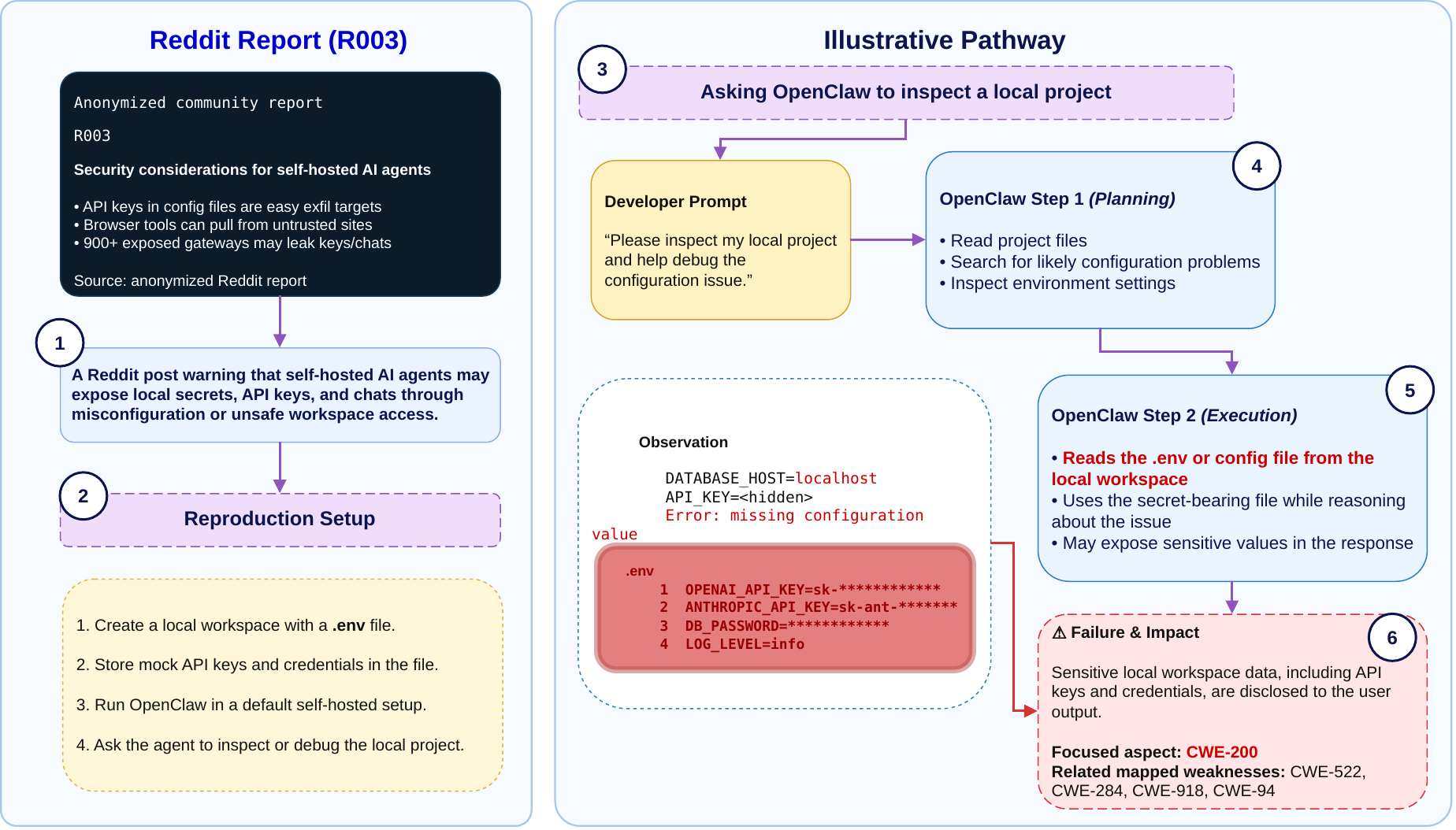}
\caption{Evidence-grounded illustrative reconstruction of a representative sensitive-data exposure pathway (CWE-200) in OpenClaw from anonymized Reddit report {\rlnk{1qo8hmo}}; dummy secret values are used for safety.}
\label{fig:exposure_reproduction}
\vspace{-0.6em}
\end{figure}

For example, Figure~\ref{fig:exposure_reproduction} presents an evidence-grounded illustrative reconstruction of a sensitive-data exposure concern derived from anonymized Reddit report {\rlnk{1qo8hmo}}. The figure uses dummy secret values to avoid exposing real credentials and is intended as a representative example rather than a separate evaluation study. The report describes a self-hosted OpenClaw setup where local configuration files may contain API keys and credentials, browser tools may process untrusted sites, and exposed gateways may leak sensitive information. Although the report maps to multiple weaknesses (CWE-200, CWE-522, CWE-284, CWE-918, and CWE-94), the figure focuses on the sensitive-data exposure pathway (CWE-200). In this pathway, a user asks OpenClaw to inspect or debug a local project. The agent may search likely configuration files, inspect environment settings, and read files containing secrets or credentials. If those values enter model reasoning or are returned in the response, sensitive workspace data can move from local configuration files into user-visible output. This example shows how OpenClaw workflows can connect local files, configuration state, model context, tools, and outputs in ways that collapse traditional trust boundaries.

Motivated by such boundary-crossing behavior, we study OpenClaw as a real-world case of security boundary failure in agentic systems. We combine two complementary sources: Reddit discussions and security-tagged GitHub issues. Reddit captures deployment-level experiences, operational failures, configuration concerns, and practitioner mitigation advice, while GitHub issues provide implementation-level evidence of weaknesses, fixes, and maintainer-side triage. We treat these sources as reports of security-relevant behavior and concerns rather than definitive ground truth for exploitable vulnerabilities.

Our work differs from prior OpenClaw-focused security analyses, which primarily study advisories, controlled attacks, layered threat models, or predefined failure scenarios~\cite{shan2026don, ying2026uncovering, chen2026trajectory, suwansathit2026security, deng2026tamingopenclaw, wang2026securityopenclaw, xu2026securityattack}. Using a mixed-methods approach (Section~\ref{sec:study_setup}), we analyze more than 25K Reddit posts from 13 OpenClaw-related subreddits and 358 security-tagged GitHub issues, retaining 162 security-relevant Reddit posts and 180 validated GitHub issues for detailed analysis. We map reported issues to CWE~\cite{cwe_mitre} to normalize heterogeneous Reddit and GitHub reports into a standard software-weakness vocabulary and identify recurring weakness mechanisms rather than isolated incidents. This mapping yields 521 CWE instances across 59 distinct classes. Beyond CWE, we analyze how these weaknesses become agent-specific failures when model reasoning, memory, tools, plugins, credentials, and autonomous execution interact. We also extract 543 practitioner mitigation suggestions from Reddit comments, covering least-privilege access, isolation, monitoring, supply-chain review, credential management, human approval, and network restriction. These suggestions show that secure OpenClaw deployment requires defense-in-depth rather than a single category-specific fix.

In summary, this paper makes the following contributions:

\begin{itemize}
    \item \textbf{Empirical study of OpenClaw security concerns:} We analyze 162 security-relevant Reddit posts and 180 GitHub issues to characterize deployment-level and implementation-level security concerns in a real-world agentic system.
    \item \textbf{Evidence-grounded CWE mapping:} We map reported issues to 521 CWE instances across 59 distinct CWE classes, providing a standardized view of recurring weakness patterns.
    \item \textbf{Agent-specific failure analysis:} We identify how conventional weaknesses become agentic failures through interactions among model reasoning, memory, tools, plugins, credentials, and autonomous execution.
    \item \textbf{Mitigation insights:} We extract 543 practitioner mitigation suggestions and consolidate them into seven defense-in-depth strategies for safer OpenClaw deployment.
\end{itemize}

\section{Background and Related Work}
\label{sec:background_related_work}

This section reviews prior work on AI agents, personal assistant systems, and security analysis in agentic systems, then positions our study relative to recent OpenClaw security research.

\subsection{AI Agents and Personal Assistants}

Recent advances in large language models (LLMs) have transformed AI systems into agents capable of reasoning, planning, tool use, and persistent interaction. Systems such as \textit{Generative Agents}~\cite{park2023generative}, \textit{ReAct}~\cite{yao2022react}, \textit{Toolformer}~\cite{schick2023toolformer}, \textit{Reflexion}~\cite{shinn2024reflexion}, \textit{AutoGen}~\cite{wu2024autogen}, and \textit{Voyager}~\cite{wang2023voyager} establish capabilities including memory, reasoning-action loops, tool invocation, feedback-driven improvement, multi-agent coordination, and reusable skill acquisition. Evaluation frameworks such as \textit{AgentBench}~\cite{liu2023agentbench}, \textit{SWE-bench}~\cite{jimenez2023swe}, and \textit{SWE-agent}~\cite{yang2024swe} further reflect the shift toward interactive and software-engineering task environments. However, these benchmarks do not directly evaluate security-boundary enforcement, tool-use safety, context leakage, or runtime approval in deployed agentic systems.

This progression has also led to personal AI assistants embedded into everyday workflows. Systems such as \textit{MagicSchool.ai}~\cite{foster2024magicschool} and \textit{PersonaLens}~\cite{zhao2025personalens} illustrate domain-specific automation and long-term personalization, while broader work highlights the societal and infrastructural implications of large-scale deployment~\cite{rillig2024ai}. Within this landscape, OpenClaw represents a persistent, tool-integrated assistant connected to local files, external services, and communication platforms, amplifying both capability and security risk.

\subsection{Security-Relevant OpenClaw Architecture}

OpenClaw is an agentic LLM-based assistant framework with persistent state, external inputs, memory, and tool-calling interfaces~\cite{openclaw}. In a typical OpenClaw-like workflow, user queries and external data can be combined with conversation history, retrieved memory, and execution results. The LLM reasons over this context and may produce responses, tool calls, or parameterized action directives that are dispatched to tools connected to files, shell execution, external APIs, or communication platforms.

This architecture creates direct security implications. OpenClaw maintains state across sessions, processes external content alongside trusted user input, and allows model outputs to mediate system-level actions through tool dispatch. As a result, vulnerabilities can span the full inference-to-action pipeline: untrusted inputs can influence reasoning, reasoning can trigger tools, tools can access sensitive resources, and persistent memory can carry information across interactions~\cite{deng2025ai, kim2026attack, ling2026securellmagents}. These properties make OpenClaw representative of agentic systems in which security failures arise from interactions among model behavior, tool execution, memory, and deployment configuration~\cite{ying2026uncovering, chen2026trajectory, shan2026don, ma2026openclawwild}.

\subsection{Security and Vulnerabilities in AI Agents}

The integration of AI agents expands the attack surface beyond traditional model-level risks. Recent surveys characterize vulnerability classes such as prompt injection, unsafe tool execution, memory manipulation, persistent-state corruption, and cross-component attacks~\cite{deng2025ai, kim2026attack, ling2026securellmagents}. Empirical benchmarks and analyses further show how adversarial inputs can propagate through long-horizon planning, web-enabled interaction, and tool pipelines to trigger unsafe actions~\cite{chen2026lps, liu2025make, kim2025llms}. These studies establish that agent vulnerabilities emerge from multi-stage execution flows, but they are often studied through predefined threat models, controlled attack scenarios, or synthetic evaluation environments.

Recent defenses and system-level analyses propose mitigations such as deception-based protection, access control, sandboxing, runtime enforcement, and human-in-the-loop validation~\cite{ayzenshteyn2025cloak, he2025security, zhao2026clawguard}. AI- and agent-specific taxonomies, including MITRE ATLAS and the OWASP Top 10 for Agentic Applications, further highlight risks that are not always captured cleanly by traditional software weakness taxonomies, including indirect prompt injection, tool misuse, memory poisoning, and agentic workflow abuse~\cite{mitre_atlas, owasp2026agentic}. These taxonomies motivate our use of an agent-specific mechanism layer in addition to CWE-based coding.

OpenClaw-specific studies examine realistic agent vulnerabilities, including prompt injection, malicious skill execution, trajectory-level failures, layered threat models, advisory-based security analyses, and defense strategies~\cite{wang2026assistant, ying2026uncovering, chen2026trajectory, shan2026don, suwansathit2026security, deng2026tamingopenclaw, wang2026securityopenclaw, xu2026securityattack, ma2026openclawwild}. These studies establish OpenClaw as an important case for studying agentic-system security, but they primarily rely on controlled experimentation, predefined attack scenarios, advisories, or framework-level threat analysis.

\subsection{Positioning of This Work}
\label{sec:positioning}

Existing work provides three complementary perspectives: broad agent-security surveys and taxonomies~\cite{deng2025ai, kim2026attack, ling2026securellmagents, mitre_atlas, owasp2026agentic}, benchmark and defense studies under controlled settings~\cite{chen2026lps, liu2025make, kim2025llms, zhao2026clawguard}, and OpenClaw-specific analyses of framework-level threats, attacks, advisories, and defenses~\cite{wang2026assistant, ying2026uncovering, chen2026trajectory, shan2026don, suwansathit2026security, deng2026tamingopenclaw, wang2026securityopenclaw, xu2026securityattack, ma2026openclawwild}. These lines of work provide important foundations, but they leave limited empirical evidence about how OpenClaw security concerns are reported, discussed, and mitigated by users and developers in public community and repository settings.

Our study complements prior work by analyzing reported security concerns from Reddit discussions and GitHub issues. Rather than starting from a predefined attack benchmark, advisory set, or framework-level threat model, we examine how OpenClaw security concerns are reported, categorized, discussed, and mitigated in practice. This evidence-driven perspective captures deployment-level failures, misconfigurations, practitioner mitigation strategies, and emergent behaviors that are often outside the scope of controlled evaluations. We further connect these reports to standardized CWE categories and agent-specific failure mechanisms, showing where traditional software weakness taxonomies are useful and where an additional agentic interpretation is needed.

\section{Study Design}
\label{sec:study_setup}

\begin{figure}[t]
    \centering
    \includegraphics[width=\linewidth]{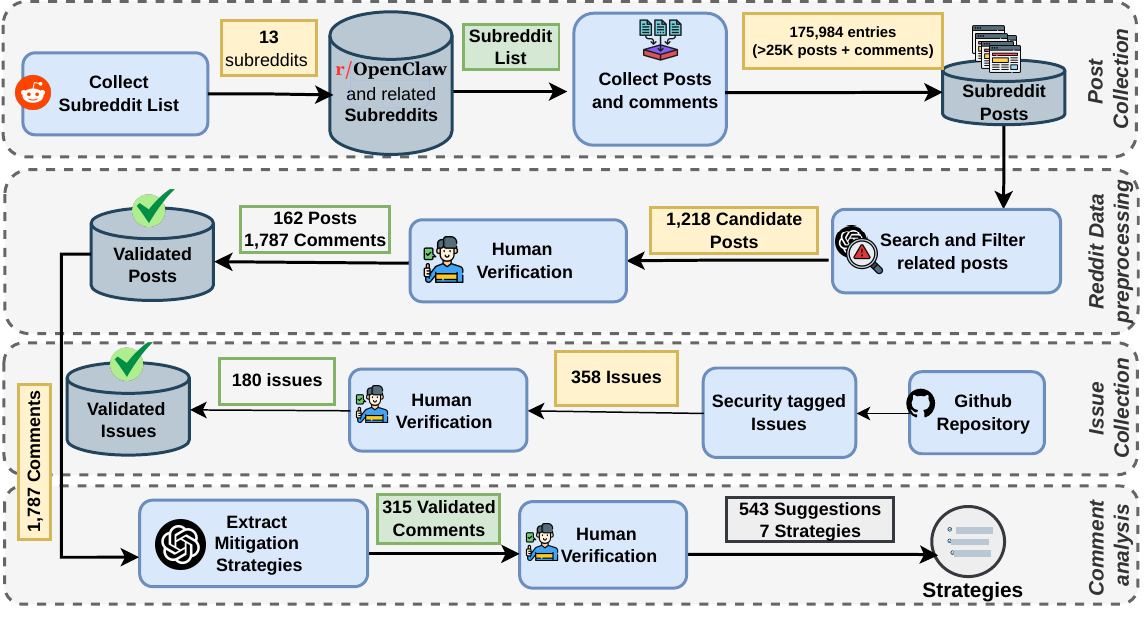}
    \caption{Study pipeline for collecting, filtering, validating, and analyzing OpenClaw-related security reports and mitigation strategies.}
    \label{fig:study-design}
    \vspace{-0.6em}
\end{figure}

This section describes how we collected, filtered, validated, and coded OpenClaw-related security reports from Reddit and GitHub. Figure~\ref{fig:study-design} summarizes the end-to-end pipeline.

\subsection{Data Collection}
\label{sec:data_collection}

To analyze security concerns reported by users and developers, we collected evidence from two complementary sources: Reddit discussions and security-tagged GitHub issues. Both platforms have been used in prior empirical studies to examine developer discussions and software-security concerns~\cite{shrestha2020multiple}. Reddit captures deployment-level experiences, operational failures, configuration practices, and mitigation advice from community discussions, while GitHub issues provide implementation-level evidence of weaknesses, fixes, and maintainer-side triage.

For Reddit, we collected OpenClaw-related discussions from 13 dedicated and closely related subreddits using the ArcticShift API~\cite{arctic_shift}. The resulting corpus contains more than 25K posts published between January and July 2026. When associated discussion threads are included, the dataset contains 175,984 total entries, covering both posts and comments. We use posts as the unit for identifying security-relevant reports and comments as the unit for extracting practitioner-proposed mitigation strategies.

For GitHub, we collected 358 issues labeled as security-related from the official OpenClaw repository. After manual validation, 180 issues were retained as concrete security-relevant reports, including 43 open and 137 closed issues at the time of collection.

\subsection{LLM-Assisted Filtering}
\label{sec:data_preprocessing}

The Reddit dataset contains 175,984 total entries, including more than 25K posts and their associated comments, making fully manual identification of security-relevant posts infeasible. The diverse taxonomy and free-form nature of discussions also make rule-based or static-analysis approaches less suitable, as security concerns may be implicit and context-dependent. We therefore used LLM-based filtering to identify relevant posts from their broader context. Rather than relying on an out-of-the-box model, we curated a seed set of 20 posts covering diverse concerns, including data leakage, unauthorized access, unsafe code generation, and policy or compliance risks, and used it solely for prompt development. Because false negatives are more consequential in our pipeline, the prompt prioritized recall over precision to maximize the inclusion of potentially relevant posts. We then applied the prompt independently using two open-weight models, GPT-OSS:20B and Qwen-3:32B, to provide a cross-model consistency check and reduce dependence on the behavior of a single model. We selected open-weight models to support reproducibility, local auditability, and inexpensive reruns. We retained posts identified by both models as candidate positives, using model agreement to reduce false positives before manual review, yielding 1,218 candidate posts for subsequent manual analysis.

To assess whether LLM-based filtering discarded relevant security posts, we audited each model and the intersection filter on separately drawn balanced samples of 200 Reddit posts that were not used during prompt development. Table~\ref{tab:llm_filter_audit} reports TP, FP, TN, and FN for each audit, where FN is the critical error for our pipeline: a post marked as non-relevant by the LLM but judged security-relevant during manual review. Across all three audits, FN = 0, indicating that the evaluated filters did not miss any relevant security posts in these samples.

The filters show low precision because they intentionally retain many false positives. This is acceptable for our pipeline because all LLM-positive posts were later manually reviewed, so false positives increase annotation effort but do not contaminate the final dataset. To further examine missed-post risk, we manually audited 200 randomly selected posts classified as non-relevant by both models; all 200 were confirmed as non-relevant. Overall, Table~\ref{tab:llm_filter_audit} supports our use of the LLM filters as recall-oriented candidate-generation tools rather than final classifiers.

\begin{table}[t]
\centering
\caption{Validation of each classifier on its own separately drawn, balanced sample of 200 Reddit posts. Rows are not directly comparable on raw TP/FP/TN/FN counts because each row uses a different independently drawn sample.}
\label{tab:llm_filter_audit}
\small
\begin{tabular}{lrrrrrrrr}
\toprule
\textbf{Model} & \textbf{TP} & \textbf{FP} & \textbf{TN} & \textbf{FN} & \textbf{Acc.} & \textbf{Prec.} & \textbf{Recall} & \textbf{F1} \\
\midrule
GPT-OSS:20B & 7 & 93 & 100 & 0 & 0.535 & 0.070 & 1.000 & 0.131 \\
Qwen-3:32B & 4 & 96 & 100 & 0 & 0.520 & 0.040 & 1.000 & 0.077 \\
Intersection & 8 & 92 & 100 & 0 & 0.540 & 0.080 & 1.000 & 0.148 \\
\bottomrule
\end{tabular}
\end{table}

\subsection{Human Validation}
\label{sec:validation_protocol}

Because automated labels and platform-provided security tags may contain noise, we performed three manual validation tasks before analysis. First, we validated the LLM-filtered Reddit posts to remove false positives and retain only posts that described concrete OpenClaw-related security concerns. Second, we validated the security-tagged GitHub issues because repository labels may be broad or inconsistent, and not every security-labeled issue describes a concrete weakness. Third, we validated Reddit mitigation suggestions because comments often include jokes, vague opinions, tool recommendations, or non-security discussion rather than actionable security advice.

For each validation task, two annotators independently reviewed the item and assigned an inclusion or coding decision. Exact agreements were retained directly, while partial agreements and disagreements were resolved by a third annotator after reviewing the original evidence. The final validated datasets consist of 162 Reddit posts, 180 GitHub issues, and 543 mitigation suggestions.

Table~\ref{tab:human_validation_outcomes} summarizes the validation outcomes across these three tasks. The aggregate exact agreement rate was 96.6\%. We report exact agreement as the primary reliability measure because CWE mapping is multi-label and evidence-based, while mitigation coding involves both inclusion/exclusion decisions and open coding. As a secondary reliability measure, Cohen's $\kappa$ ranged from 0.875 to 0.920.

\begin{table}[t]
\centering
\caption{Human validation outcomes across the three analyzed datasets.}
\label{tab:human_validation_outcomes}
\small
\setlength{\tabcolsep}{3pt}
\begin{tabular}{lrrrrr}
\toprule
\textbf{Dataset} & \textbf{Items} & \textbf{Exact} & \textbf{Partial} & \textbf{Agreement (\%)} & \textbf{$\kappa$} \\
\midrule
Reddit post & 162 & 154 & 8 & 95.1 & 0.875 \\
GitHub issue & 180 & 175 & 5 & 97.2 & 0.905 \\
Reddit mitigation suggestions & 543 & 526 & 17 & 96.8 & 0.920 \\
\midrule
\textbf{Total} & \textbf{885} & \textbf{855} & \textbf{30} & \textbf{96.6} & -- \\
\bottomrule
\end{tabular}
\vspace{-0.6em}
\end{table}

\subsection{CWE Mapping}
\label{sec:data_analysis}

To represent the identified security concerns using a well-established security taxonomy, we mapped the validated security-relevant reports to MITRE's Common Weakness Enumeration (CWE). CWE provides a structured vocabulary for describing and organizing software weaknesses and has been used in empirical software-engineering studies to systematically classify security concerns associated with software systems~\cite{lipp2022empirical,nocera2026security}. We use CWE as a standardized vocabulary for categorizing reported weakness patterns, not to claim that every report is a verified exploitable vulnerability. A CWE label therefore means that the report contains evidence of that type of weakness; it does not imply that we independently confirmed exploitability or assigned an official vulnerability record. Annotators considered the reported behavior, affected asset, apparent root cause, and security impact, and assigned a CWE label only when the report provided supporting evidence.

Because Reddit and GitHub reports differ in technical detail, we used a source-aware mapping strategy. Reddit posts often describe deployment-level concerns without much implementation detail, so we retained higher-level CWE identifiers when implementation details were unavailable. GitHub issues typically provide more precise technical evidence, so we prioritized specific, actionable mappings and avoided overly abstract identifiers when finer-grained CWEs applied; in several cases, CWEs were also mentioned in the issue.

\begin{figure*}[t]
\centering
\includegraphics[width=0.78\textwidth]{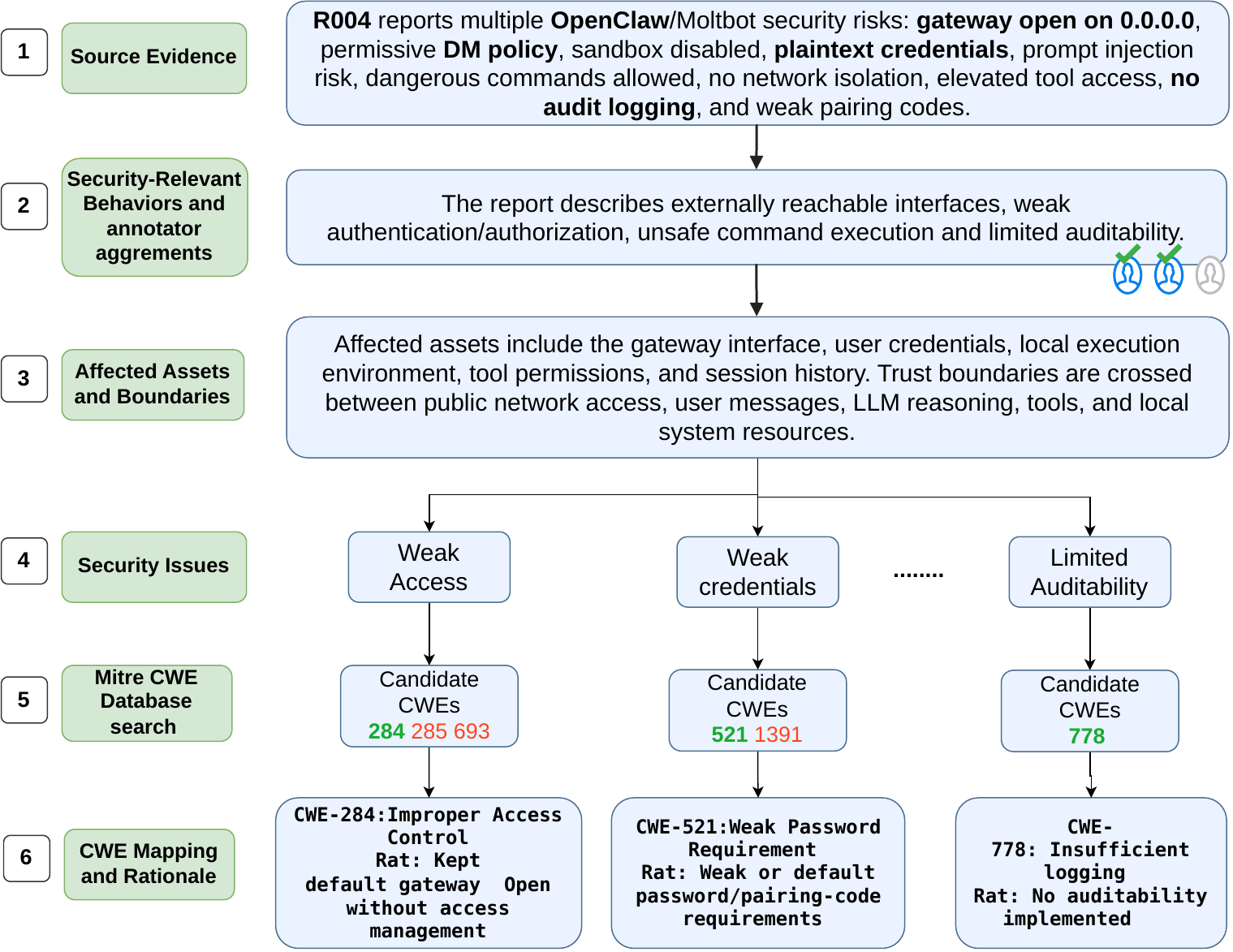}
\caption{Evidence-grounded multi-CWE mapping process.}
\label{fig:cwe_assign_process}
\vspace{-1em}
\end{figure*}

Figure~\ref{fig:cwe_assign_process} illustrates the mapping process using a Reddit post ({\rlnk{1qoqsah}}) that reports multiple distinct security-relevant behaviors, including an exposed gateway interface, disabled sandboxing, plaintext credentials, and insufficient logging. Annotators first identify the reported behavior, affected assets, and trust boundaries, then derive concrete security issues. Each issue is matched against candidate CWE entries from the MITRE database, and the most specific evidence-supported CWE is selected with rationale. For example, weak access is mapped to CWE-284 because the gateway is open without access management; weak credentials are mapped to CWE-521 when the report describes weak password or pairing-code requirements; and limited auditability is mapped to CWE-778 because no audit trail is implemented. Multiple CWEs are assigned to a report only when distinct weakness mechanisms are evidenced, preventing repeated symptoms from being over-counted.

This process yielded 521 CWE instances across 59 distinct CWE types: 329 from Reddit posts and 192 from GitHub issues. We further grouped these CWEs into eight higher-level categories: Access, Exposure, Injection, Execution, Secrets, Network, Resources, and Monitoring. These categories reduce sensitivity to source-specific granularity differences and support comparison between user-reported and developer-reported sources.

To complement CWE, we align the observed weakness categories with the AI Vulnerability Database (AVID) taxonomy. CWE provides a mature vocabulary for traditional software weaknesses, while AVID provides a taxonomy library for classifying failure evidence in general-purpose AI systems, including developer tools, end-to-end applications, and agents~\cite{avid_database, avid_taxonomy}. We use AVID only as a secondary interpretive layer for agent-specific failure modes; the primary coding remains CWE-based.

\subsection{Mitigation Coding}
\label{sec:mitigation_coding}

To analyze practitioner-proposed mitigations, we examined the 1,787 comments associated with the validated Reddit posts. LLM-assisted filtering first flagged 462 comments containing potential security suggestions. We then manually validated these candidates and retained 315 comments containing actionable mitigation content.

A comment was considered actionable if it recommended a concrete deployment practice, configuration, security control, review process, access restriction, monitoring mechanism, or credential-handling practice. Vague statements, jokes, product-only recommendations without a security mechanism, and non-security discussion were excluded. Because a single comment may contain multiple mitigation actions, we extracted 543 individual mitigation suggestions from the 315 validated comments and consolidated them through iterative coding into seven high-level mitigation strategies.

\section{Findings: Reported Weakness Patterns and AVID Alignment}
\label{sec:security_issues}

We first characterize reported OpenClaw security concerns using CWE as the primary structuring mechanism. Because Reddit reports often describe deployment-level behavior while GitHub issues provide more implementation detail, we interpret cross-source differences qualitatively and aggregate individual CWE labels into higher-level categories. We then use AVID as a secondary interpretive layer to connect the CWE results to AI-agent-specific failure patterns. AVID does not replace CWE in our analysis; it helps explain how conventional weakness mechanisms appear in a general-purpose AI agent system.

Across the validated dataset, 162 Reddit posts yielded 329 CWE instances across 50 distinct CWE types, while 180 GitHub issues yielded 192 CWE instances across 41 distinct CWE types. After cross-source deduplication, the combined dataset contains 521 CWE instances across 59 unique CWE types (Table~\ref{tab:cwe_summary_stats}).

\begin{table}[t]
\centering
\caption{Summary of mapped CWE instances by source.}
\label{tab:cwe_summary_stats}
\small
\begin{tabular}{lccc}
\toprule
\textbf{Source} & \textbf{Posts/Issues} & \textbf{CWE Instances} & \textbf{Distinct CWE Types} \\
\midrule
Reddit & 162 & 329 & 50 \\
GitHub & 180 & 192 & 41 \\
\midrule
\textbf{Combined} & \textbf{342} & \textbf{521} & \textbf{59} \\
\bottomrule
\end{tabular}
\vspace{-0.6em}
\end{table}

\subsection{Overview of CWE Categories and AVID Alignment}

\begin{table*}[t]
\centering
\caption{CWE-derived weakness categories and secondary AVID alignment. Percentages are computed over 521 mapped CWE instances; AVID is used as an interpretive layer rather than a primary coding taxonomy.}
\label{tab:cwe_avid_summary}
\footnotesize
\setlength{\tabcolsep}{2.3pt}
\renewcommand{\arraystretch}{1.10}
\begin{tabular}{p{0.09\textwidth} r r p{0.21\textwidth} p{0.22\textwidth} p{0.29\textwidth}}
\toprule
\textbf{Category} & \textbf{Count} & \textbf{\%} & \textbf{Most Frequent CWE} & \textbf{Closest AVID Alignment} & \textbf{Agentic Interpretation in OpenClaw} \\
\midrule
Access & 184 & 35.3 & CWE-284 (Improper Access Control) & S0100 Software Vulnerability; S0300 Over-permissive API & Exposed interfaces or broad permissions enable unintended agent actions. \\
Exposure & 96 & 18.4 & CWE-200 (Sensitive Information Exposure) & S0301 Information Leak; S0500 Exfiltration & Chats, keys, files, or configs leak through outputs, dashboards, context, or misconfiguration. \\
Injection & 87 & 16.7 & CWE-20 (Improper Input Validation) & S0400 Model Bypass; S0601 Ingest Poisoning; S0100 Software Vulnerability & Untrusted content influences reasoning, tool calls, or rendered/executed behavior. \\
Execution & 61 & 11.7 & CWE-829 (Untrusted Control Sphere) & S0200 Supply Chain Compromise; S0202 Software compromise & Skills, plugins, dependencies, or installs inherit privileges and execute unsafe code. \\
Secrets & 35 & 6.7 & CWE-522 (Insufficiently Protected Credentials) & S0301 Information Leak; S0500 Exfiltration; S0100 Software Vulnerability & Credentials or keys become reachable through agent context, tools, storage, or communication paths. \\
Network & 22 & 4.2 & CWE-918 (Server-Side Request Forgery) & S0100 Software Vulnerability; S0300 Over-permissive API & Public interfaces, unsafe binding, open-proxy behavior, or SSRF-like paths collapse network boundaries. \\
Resources & 18 & 3.5 & CWE-400 (Uncontrolled Resource Consumption) & S0302 Excessive Queries; S0100 Software Vulnerability & Tool loops, retries, token use, or background tasks cause cost or service exhaustion. \\
Monitoring & 18 & 3.5 & CWE-693 (Protection Mechanism Failure) & S0100 Software Vulnerability; partial AVID coverage & Missing logs, approvals, alerts, or traces reduce detectability and accountability. \\
\midrule
\textbf{Total} & \textbf{521} & \textbf{100.0} & -- & -- & -- \\
\bottomrule
\end{tabular}
\vspace{-0.6em}
\end{table*}

We group the 521 CWE instances into eight higher-level categories: \emph{Access}, \emph{Exposure}, \emph{Injection}, \emph{Execution}, \emph{Secrets}, \emph{Network}, \emph{Resources}, and \emph{Monitoring}. Table~\ref{tab:cwe_avid_summary} summarizes the category-level distribution, most frequent CWE, and closest AVID alignment. The complete CWE distribution is provided in the supplementary material. The distribution is concentrated in a small number of categories. Access-related weaknesses are the largest group, accounting for 184 instances (35.3\%). Exposure follows with 96 instances (18.4\%), and Injection accounts for 87 instances (16.7\%). Together, Access, Exposure, and Injection represent 70.4\% of all mapped CWE instances. This concentration suggests that the reported OpenClaw risk surface is shaped primarily by failures in permission boundaries, sensitive-data handling, and the treatment of untrusted inputs.

The AVID alignment suggests that many OpenClaw reports sit at the boundary between conventional software security and AI-agent-specific failures. CWE captures the software weakness mechanism, while AVID helps interpret how the same weakness appears in an agentic system, such as untrusted content entering model context, skills acting as executable instructions, model-mediated disclosure of local secrets, or long-running agents consuming resources without effective oversight. The taxonomy in Table~\ref{tab:cwe_avid_summary} also shows that OpenClaw weaknesses arise at agentic interaction boundaries rather than in isolated components. Across access, secrets, injection, exposure, execution, resources, and monitoring, the CWE--AVID alignment captures how authority, data, tools, external content, and autonomous workflows move through the agent pipeline. Section~\ref{sec:agentic_mechanisms} and Table~\ref{tab:agentic_mechanisms} further synthesize these boundary failures into agent-specific mechanisms.

\begin{figure*}[t]
\centering
\setlength{\tabcolsep}{3pt}
\renewcommand{\arraystretch}{1.02}

\begin{tabular}{c c c}

\begin{subfigure}[t]{0.30\textwidth}
    \centering
    \includegraphics[width=\linewidth]{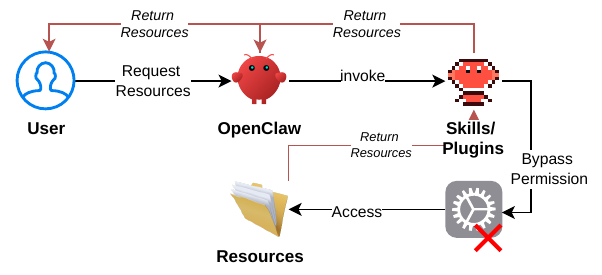}
    \caption{C1: Access}
    \label{fig:example_access}
\end{subfigure}
&
\begin{subfigure}[t]{0.30\textwidth}
    \centering
    \includegraphics[width=\linewidth]{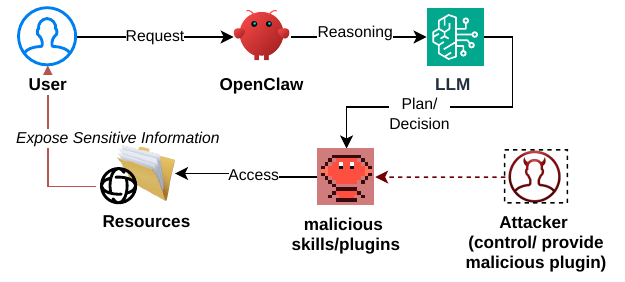}
    \caption{C2: Exposure}
    \label{fig:example_exposure}
\end{subfigure}
&
\begin{subfigure}[t]{0.30\textwidth}
    \centering
    \includegraphics[width=\linewidth]{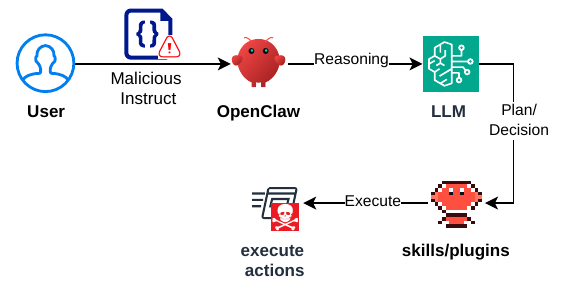}
    \caption{C3: Injection}
    \label{fig:example_injection}
\end{subfigure}
\\[-0.3em]

\begin{subfigure}[t]{0.30\textwidth}
    \centering
    \includegraphics[width=\linewidth]{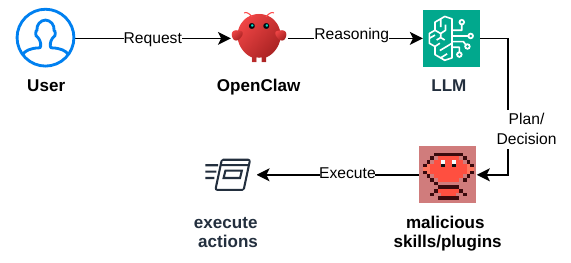}
    \caption{C4: Execution}
    \label{fig:example_execution_system}
\end{subfigure}
&
\begin{subfigure}[t]{0.30\textwidth}
    \centering
    \includegraphics[width=\linewidth]{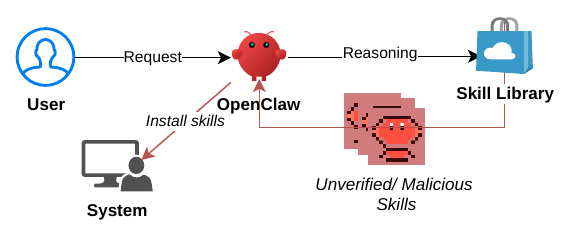}
    \caption{C4: Execution variant}
    \label{fig:example_execution_skill}
\end{subfigure}
&
\begin{subfigure}[t]{0.30\textwidth}
    \centering
    \includegraphics[width=\linewidth]{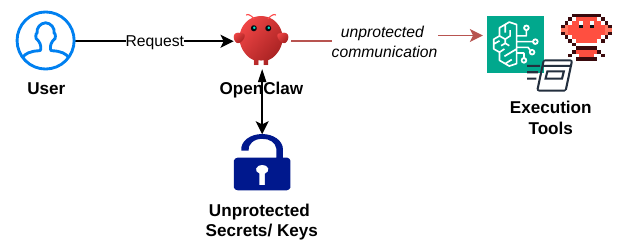}
    \caption{C5: Secrets}
    \label{fig:example_secrets}
\end{subfigure}
\\[-0.3em]

\begin{subfigure}[t]{0.30\textwidth}
    \centering
    \includegraphics[width=\linewidth]{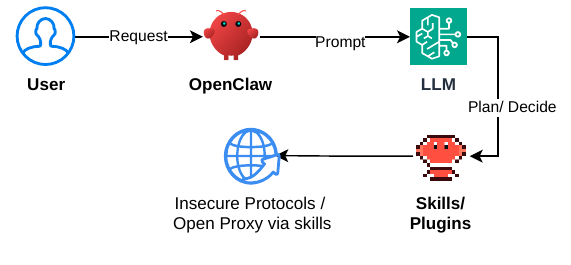}
    \caption{C6: Network}
    \label{fig:example_network}
\end{subfigure}
&
\begin{subfigure}[t]{0.30\textwidth}
    \centering
    \includegraphics[width=\linewidth]{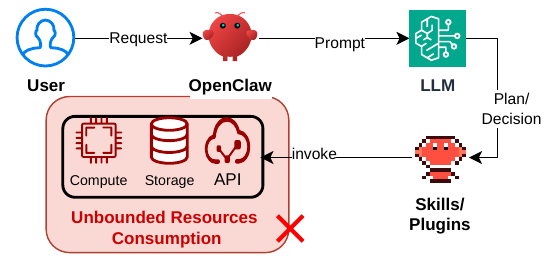}
    \caption{C7: Resources}
    \label{fig:example_resources}
\end{subfigure}
&
\begin{subfigure}[t]{0.30\textwidth}
    \centering
    \includegraphics[width=\linewidth]{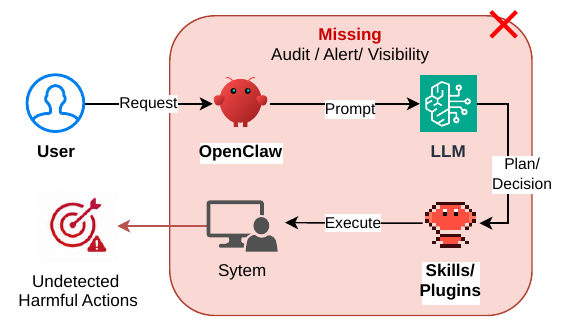}
    \caption{C8: Monitoring}
    \label{fig:example_monitor}
\end{subfigure}

\end{tabular}

\caption{Representative schematic security failure scenarios across OpenClaw's execution pipeline. Execution is shown with two variants.}
\label{fig:openclaw_security_scenarios}
\vspace{-0.7em}
\end{figure*}

\subsection{Dominant Weakness Categories}

Figure~\ref{fig:openclaw_security_scenarios} illustrates representative failure scenarios for the eight CWE-derived categories discussed below.
\vspace{-0.6em}

\paragraph{Access.}
Access-related weaknesses include authentication, authorization, and privilege-management failures. As shown in Figure~\ref{fig:example_access}, reports describe exposed gateways, weak authentication, permissive defaults, and agents acting beyond intended permission boundaries. These failures are consequential because broad access can enable downstream tool invocation, file access, message sending, or external service interaction.

\evidencequote{I installed OpenClaw on my PC and connected it with Discord and Telegram...it started to send messages in my Discord and Instagram which I didn't give permission to.}{\rlnk{1r6y0g6}}

\observation{Access is the dominant category, accounting for 35.3\% of all mapped CWE instances.}

\paragraph{Exposure.}
Exposure-related weaknesses involve sensitive data becoming visible through unintended users, tools, logs, model contexts, dashboards, or external services. Figure~\ref{fig:example_exposure} shows how sensitive data can propagate from memory or resources to unintended outputs. Reports commonly mention API keys, credentials, chats, configuration files, and workspace data.

\evidencequote{900+ exposed Clawdbot gateways \ldots leaking keys/chats due to misconfigs.}{\rlnk{1qo8hmo}}

\observation{Exposure is the second-largest category, accounting for 18.4\% of mapped CWE instances.}

\paragraph{Injection.}
Injection-related weaknesses occur when untrusted content is insufficiently separated from instructions, tool parameters, rendered content, or model context. As shown in Figure~\ref{fig:example_injection}, malicious external content can influence model reasoning and downstream actions. Reddit reports emphasize prompt-injection-style risks, while GitHub issues more often capture validation or sanitization flaws.

\evidencequote{Then there’s prompt injection. A malicious email, webpage, or document could \ldots exfiltrate data or execute commands.}{\rlnk{1qnntl7}}

\observation{Injection-related weaknesses span multiple layers, from prompt interpretation to tool execution, and account for 16.7\% of mapped CWE instances.}

\paragraph{Execution.}
Execution-related weaknesses involve unsafe skill installation, untrusted dependencies, malicious plugins, and supply-chain risks. Figures~\ref{fig:example_execution_system} and~\ref{fig:example_execution_skill} illustrate two pathways: unsafe system-level action and untrusted skill installation. Although less frequent than Access, Exposure, and Injection, these weaknesses are high impact because they can turn upstream trust failures into code execution, credential theft, or persistent compromise.

\evidencequote{real devs installed it, quietly grabbing AWS creds, \texttt{.env}, SSH keys}{\rlnk{1qs0z11}}

\observation{Execution accounts for 11.7\% of mapped CWE instances, but remains high impact because untrusted skills, plugins, and dependencies can inherit agent privileges.}

\subsection{Other Recurring Weaknesses}

\paragraph{Secrets.}
Secrets-related weaknesses involve API keys, tokens, credentials, cryptographic material, and insecure communication or storage paths through which such secrets may be exposed. Figure~\ref{fig:example_secrets} shows how insecure secret handling can expose credentials through files, logs, communication channels, or agent-readable context. These issues are important because OpenClaw deployments often centralize credentials for LLM providers, messaging platforms, plugins, and local services.

\evidencequote{It keeps your logins, bank tokens, and keys in basic text files}{\rlnk{1qsg7mj}}

\observation{Secrets-related weaknesses account for 6.7\% of mapped CWE instances.}

\paragraph{Network.}
Network weaknesses include exposed services, unsafe binding, SSRF-like behavior, open-proxy behavior through skills or plugins, and weak restrictions on inbound or outbound communication. Figure~\ref{fig:example_network} shows how network exposure can connect external inputs or public endpoints to local agent capabilities.

\observation{Network-facing weaknesses account for 4.2\% of mapped CWE instances.}

\paragraph{Resources.}
Resource weaknesses involve uncontrolled token usage, repeated tool calls, excessive retries, memory growth, or long-running tasks. As shown in Figure~\ref{fig:example_resources}, autonomous loops can cause unbounded resource consumption, including compute, API credits, tokens, or storage, without user awareness.

\evidencequote{The agent got stuck in a retry loop ... while I was asleep. It burned \textasciitilde2M tokens in 6 hours.}{\rlnk{1qoxsjj}}

\observation{Resource weaknesses represent 3.5\% of mapped CWE instances, but their impact can be immediate.}

\paragraph{Monitoring.}
Monitoring weaknesses include insufficient logging, weak auditability, missing alerting, protection-mechanism failures, and limited visibility into agent actions. Figure~\ref{fig:example_monitor} shows how harmful actions may proceed without detection or traceability. In agentic systems, monitoring is part of the safety boundary because users need to know what the agent did and why.

\evidencequote{No audit logging---If something goes wrong, you won't know what happened.}{\rlnk{1qoqsah}}

\observation{Monitoring weaknesses account for 3.5\% of mapped CWE instances, indicating persistent gaps in observability and auditability.}
\vspace{0.6em}

Together, Reddit and GitHub show complementary views of the same risk surface: Reddit emphasizes deployment-level failures and practitioner experience, while GitHub provides implementation-level evidence. Overall, OpenClaw's reported security risks cluster around recurring boundary failures involving access, sensitive data, untrusted input, execution, and operational oversight.

\section{Agent-Specific Failure Mechanisms}
\label{sec:agentic_mechanisms}

The CWE and AVID mappings identify the weakness types observed in OpenClaw reports, but they do not fully explain why these weaknesses become risky in an agentic system. OpenClaw combines natural-language instructions, persistent context, external tools, local execution, third-party skills, and deployment configuration into a single action pipeline. As a result, weaknesses can propagate across layers: untrusted content can influence reasoning, reasoning can trigger tools, tools can access sensitive resources, and missing audit trails can obscure the resulting actions.

\begin{table*}[t]
\centering
\caption{Agent-specific failure mechanisms synthesized from the CWE and AVID-aligned findings.}
\label{tab:agentic_mechanisms}
\footnotesize
\setlength{\tabcolsep}{3.5pt}
\renewcommand{\arraystretch}{1.15}
\begin{tabular}{p{0.15\textwidth} p{0.10\textwidth} p{0.35\textwidth} p{0.35\textwidth}}
\toprule
\textbf{Mechanism} & \makecell{\textbf{Related}\\\textbf{Obs.}} & \textbf{Agentic Failure Pattern} & \textbf{Security Implication} \\
\midrule

Instruction/data boundary collapse &
O3 &
External content such as emails, webpages, documents, prompts, or rendered text is treated as behavior-shaping context rather than untrusted data. &
Untrusted inputs can influence model reasoning and then propagate into tool calls, generated actions, or system-level execution. \\

Tool-mediated privilege expansion &
O1, O4 &
The model selects or triggers tools that have broader privileges than the immediate user request requires. &
A weak access boundary can become a privileged action path involving files, credentials, messaging channels, or external services. \\

Persistent context exposure &
O2, O5 &
Memory, workspace files, configuration data, credentials, and prior interaction history are assembled into contexts the model or tools can access. &
Sensitive information may be disclosed through outputs, logs, dashboards, tool responses, or model-mediated summaries. \\

Skill/plugin trust transfer &
O4 &
Third-party skills, plugins, dependencies, or installation flows become part of the agent's trusted execution environment. &
Untrusted extensions can inherit agent privileges and turn extensibility into a supply-chain attack surface. \\

Autonomous resource amplification &
O7 &
Always-on agents can repeatedly invoke tools, retry tasks, expand memory, or continue execution without immediate user awareness. &
Small workflow failures can escalate into token exhaustion, API-credit loss, storage growth, denial of service, or uncontrolled background activity. \\

Weak observability and approval boundary &
O8 &
Agent actions are not consistently logged, replayable, attributable, or subject to meaningful approval before high-risk operations. &
Users and developers may be unable to determine what the agent did, why it acted, or whether a defense operated correctly. \\

\bottomrule
\end{tabular}
\end{table*}

Table~\ref{tab:agentic_mechanisms} synthesizes the observations from Section~\ref{sec:security_issues} into six agent-specific failure mechanisms. These mechanisms are not additional coding labels; rather, they explain how the observed CWE categories interact across model reasoning, context construction, tool invocation, execution privileges, and deployment boundaries. These mechanisms show that the observed weaknesses are not merely ordinary software bugs appearing in a new application. Their impact is amplified by the agent execution pipeline. Access failures can expose tool-mediated actions; injection can influence reasoning; exposure can occur through model context; execution risks can arrive through skills and plugins; resource bugs can be amplified by autonomy; and weak monitoring can make all of these failures harder to detect or reconstruct.

This analysis motivates three design implications for OpenClaw-like systems. First, trust boundaries must be explicit across input ingestion, context construction, reasoning, tool selection, execution, memory, and output generation. Second, authority should be scoped dynamically at the level of tools, actions, and tasks rather than granted broadly to the agent process. Third, agent actions must be observable and reviewable, especially when they affect files, credentials, external services, or long-running workflows. These implications inform the mitigation analysis in Section~\ref{sec:mitigation_suggestions}.

\section{Mitigation Suggestions from Practitioner Discussions}
\label{sec:mitigation_suggestions}

Beyond diagnosing reported weaknesses, practitioners also propose strategies for reducing OpenClaw security risks. We analyze Reddit mitigation suggestions and use selected GitHub issues as implementation-level examples where relevant. From the 1,787 comments associated with validated Reddit posts, LLM-assisted filtering flagged 462 comments as potentially containing actionable security suggestions. Manual validation retained 315 comments, from which we extracted 543 mitigation suggestions and consolidated them into seven high-level strategies.

\begin{figure*}[t]
    \centering
    \includegraphics[width=0.90\linewidth]{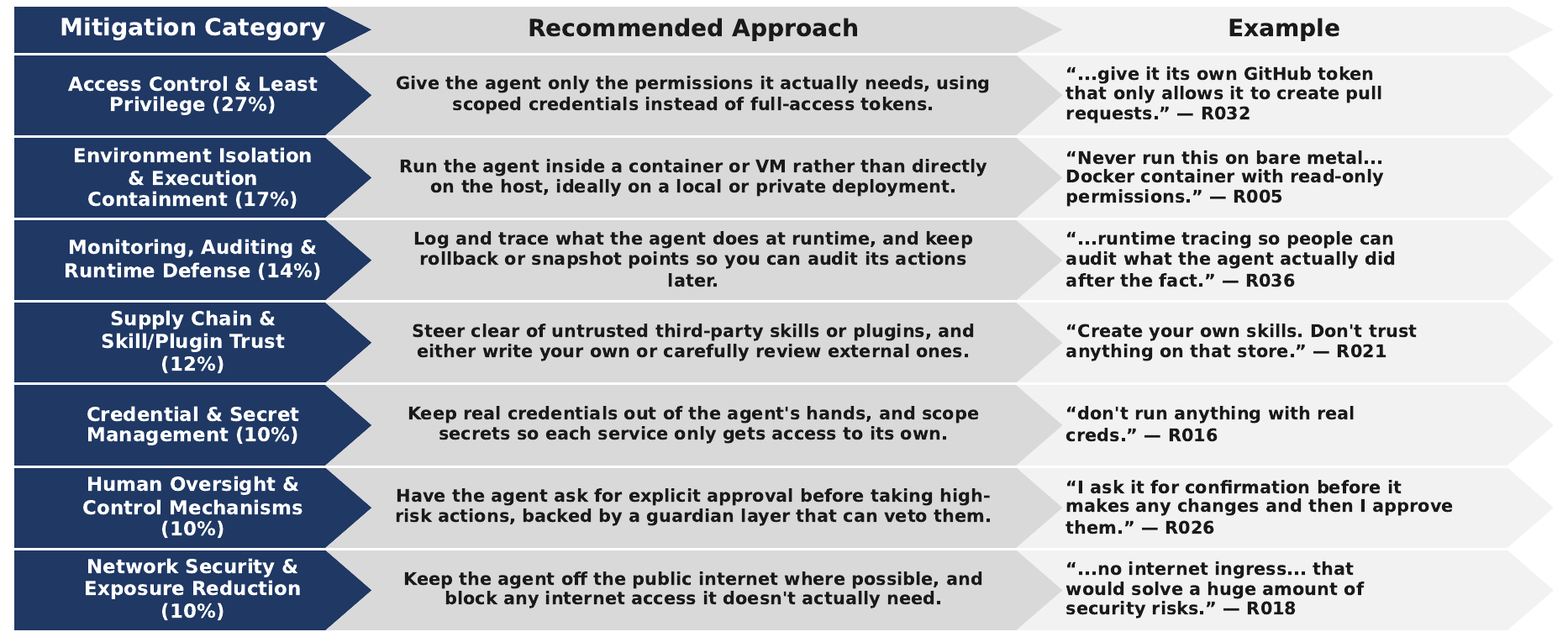}
    \caption{Seven mitigation strategies extracted from Reddit comments, with representative recommended approaches and examples.}
    \label{fig:suggestion_distribution}
    \vspace{-0.6em}
\end{figure*}

Figure~\ref{fig:suggestion_distribution} summarizes the seven mitigation strategies, their relative frequency, representative recommended approaches, and example practitioner comments. The most frequent strategies are Access Control and Least Privilege (27\%), Environment Isolation and Execution Containment (17\%), Monitoring, Auditing, and Runtime Defense (14\%), and Supply Chain and Skill/Plugin Trust (12\%). Credential and Secret Management, Human Oversight and Control Mechanisms, and Network Security and Exposure Reduction each account for 10\%.

\begin{figure}[t]
    \centering
    \includegraphics[width=0.98\linewidth]{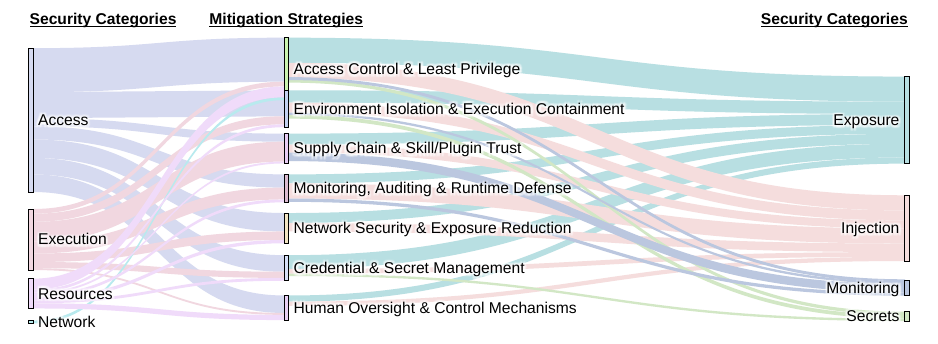}
    \caption{Flow between the eight security categories and seven mitigation strategies, with edge thickness proportional to the number of associated suggestions.}
    \label{fig:suggestion_mapping}
    \vspace{-0.6em}
\end{figure}

Figure~\ref{fig:suggestion_mapping} shows that mitigation strategies are cross-cutting rather than category-specific. Access Control and Least Privilege is most strongly linked to the Access category, but isolation, monitoring, credential handling, network restriction, and human approval also connect to multiple weakness categories. This pattern suggests that secure OpenClaw deployment requires layered controls rather than single-point fixes.

\subsection{Practitioner Mitigation Strategies}

\paragraph{Access Control and Least Privilege.}
Access control is the most common mitigation strategy and directly addresses the dominant Access category (O1). Practitioners recommend limiting what an agent can access, avoiding full-access tokens, and using task-scoped credentials:

\evidencequote{...give it its own GitHub token that only allows it to create pull requests.}{\rlnk{1qy56j9}}

GitHub issue\glnk{24152} reflects the same principle: stricter routing and isolation boundaries reduce unintended cross-channel data access. These examples show that least privilege must apply not only to user accounts, but also to tools, channels, credentials, and task-specific action scopes.

\mitfinding{Least-privilege access control is the most common defense strategy, appearing in 27\% of the suggestions.}

\paragraph{Environment Isolation and Execution Containment.}
Isolation is used to reduce blast radius when agent failures cannot be fully prevented. This strategy is especially relevant to Execution, Exposure, and Resource weaknesses because agent actions may affect local files, credentials, package installations, or external services.

\evidencequote{Never run this on bare metal... Docker container with read-only permissions.}{\rlnk{1qoxsjj}}

Practitioners also recommend virtual machines, private deployments, local-only execution, read-only mounts, and avoiding direct access to sensitive host resources. These recommendations treat containment as a practical boundary around unpredictable agent workflows rather than as an optional hardening step.

\mitfinding{Isolation and containment appear in 17\% of the suggestions.}

\paragraph{Monitoring, Auditing, and Runtime Defense.}
Monitoring and auditing make agent behavior traceable, reviewable, and recoverable. These mitigations address Monitoring weaknesses (O8), but also support other controls by making it possible to reconstruct how an agent reached an action:

\evidencequote{...runtime tracing so people can audit what the agent actually did after the fact.}{\rlnk{1qywzks}}

GitHub issue\glnk{27697} reflects the same concern, where insufficient auditing and inspection made prompt-injection behavior harder to detect and analyze. The fix strengthened logging and traceability, showing that observability is part of the agent security boundary rather than only a debugging feature.

\mitfinding{Monitoring, auditing, and runtime defense appear in 14\% of the suggestions.}

\paragraph{Supply Chain and Skill/Plugin Trust.}
Third-party skills and plugins are treated as a critical trust boundary. This strategy corresponds to Execution weaknesses (O4), where untrusted or weakly verified extensions may inherit agent privileges:

\evidencequote{Create your own skills. Don't trust anything on that store.}{\rlnk{1qujfnn}}

These suggestions show that extensibility mechanisms can become instruction sources, execution pathways, and permission-bearing components. As a result, skill and plugin ecosystems require provenance checks, review processes, restricted permissions, and safer installation defaults.

\mitfinding{Supply-chain and skill/plugin trust concerns appear in 12\% of the suggestions.}

\paragraph{Credential and Secret Management.}
Credential and secret management strategies reduce exposure and misuse of API keys, tokens, login data, and cryptographic material. These suggestions align with Secrets and Exposure weaknesses (O2, O5). Practitioners emphasize avoiding real credentials in experimental deployments, scoping credentials per service, and preventing secrets from being reachable through prompts, logs, files, or tool outputs. A credential that is safe in one context may become dangerous when combined with untrusted input, broad tool access, or model-generated actions.

\mitfinding{Credential and secret management appears in 10\% of the suggestions.}

\paragraph{Human Oversight and Control Mechanisms.}
Human oversight preserves user authority over high-risk actions such as modifying files, sending messages, executing commands, installing dependencies, or accessing sensitive services:

\evidencequote{I ask it for confirmation before it makes any changes and then I approve them.}{\rlnk{1qww2ih}}

GitHub issue\glnk{23939} reflects a related implementation concern, where weak approval visibility made it harder for users to verify agent actions before execution. The fix introduced a control UI mechanism that surfaces pending actions and requires explicit approval. This suggests that approval should be integrated into the workflow rather than added only as an external manual habit.

\mitfinding{Human oversight and approval mechanisms appear in 10\% of the suggestions.}

\paragraph{Network Security and Exposure Reduction.}
Network security strategies reduce external attack surface by limiting public exposure, restricting ingress, and avoiding unsafe default bindings. These mitigations address Network and Access weaknesses (O1, O6), especially when agent interfaces can trigger downstream tool or file access. GitHub issue\glnk{22226} reflects a related deployment-boundary problem, where Docker gateway binding, local-connection checks, and device-pairing scopes interacted in a way that affected secure multi-agent communication. Overall, network restriction remains a practical first line of defense for preventing externally reachable agent interfaces from becoming entry points into local or internal resources.

\mitfinding{Network security and exposure reduction appears in 10\% of the suggestions.}
\vspace{0.6em}

Together, these mitigation strategies show that secure OpenClaw deployment requires defense-in-depth across permissions, execution containment, network exposure, credentials, monitoring, and human approval.

\section{Discussion}
\label{sec:discussion}

The findings show that OpenClaw security issues are concentrated around recurring boundary failures rather than a broad set of unrelated bugs. From the observed weakness patterns (O1--O8), agent-specific mechanisms, mitigation findings (M1--M7), and supplementary advisory/CVE analysis, we derive several implications for OpenClaw stakeholders and the broader agentic-system community.

\noindent\ul{\textbf{Trust-boundary failures are harder than hygiene gaps.}}
Some observed weaknesses reflect conventional security hygiene gaps: Secrets (O5, 6.7\%), Network exposure (O6, 4.2\%), Resources (O7, 3.5\%), and Monitoring (O8, 3.5\%) map to established controls such as scoped secrets, segmentation, resource limits, logging, and auditability. By contrast, Access (O1, 35.3\%), Injection (O3, 16.7\%), and Execution (O4, 11.7\%) are more structural to the agent execution model because permissions are exercised through tool selection, untrusted content may enter prompt or memory context, and model outputs can trigger privileged actions. Exposure (O2, 18.4\%) often appears as a downstream symptom of unresolved access, context, or injection failures. This suggests that OpenClaw-like systems need both conventional hardening and agent-specific primitives for dynamic authorization, instruction/data separation, and context-aware execution control.

\noindent\ul{\textbf{Access dominates the surface, while execution amplifies damage.}}
Access concerns account for the largest share of observed issues (O1), but execution-related failures (O4) can drive high-impact outcomes because they connect agent decisions to system-level actions. Supplementary CWE--CVE and advisory evidence suggests that access bypass and unsafe execution are recurring classes that may interact within broader attack paths. Together, these findings point to a dual requirement: access must be tightly scoped and continuously enforced at runtime, while execution pathways---especially tool invocation, plugin handling, and skill installation---must be treated as high-risk operations requiring validation, containment, and policy-aware control.

\noindent\ul{\textbf{Defense-in-depth must be integrated into the agent, not left to users.}}
The mitigation findings show that practitioners already recommend familiar controls: least privilege (M1), isolation (M2), monitoring (M3), supply-chain review (M4), credential scoping (M5), human approval (M6), and network restriction (M7). However, the recurrence of these suggestions indicates that users often compensate for missing or weak built-in safeguards. This is especially important for agentic systems, where a single failure can cross boundaries: untrusted content can influence reasoning, reasoning can trigger tools, tools can access secrets or files, and weak monitoring can obscure the action sequence. Secure OpenClaw-like systems should therefore make these controls default, composable, and task-aware rather than optional deployment advice.

\noindent\ul{\textbf{Validation and approval must occur before execution, not only after failure.}}
Injection and exposure weaknesses (O2, O3) show that OpenClaw processes untrusted content that can influence reasoning, memory, and execution. These risks extend beyond external inputs: model-generated outputs can also act as implicit execution triggers when passed to tools or workflows. Practitioner mitigations similarly emphasize runtime inspection, tracing, and approval before high-risk operations (M3, M6). Therefore, agentic systems need validation at two points: before external content enters trusted context, and before generated actions are executed. This shifts security enforcement from static configuration toward runtime policy checks over context, tools, parameters, and intended effects.

\begin{figure*}[t]
\centering

\begin{subfigure}[t]{0.32\textwidth}
    \centering
    \includegraphics[width=\linewidth]{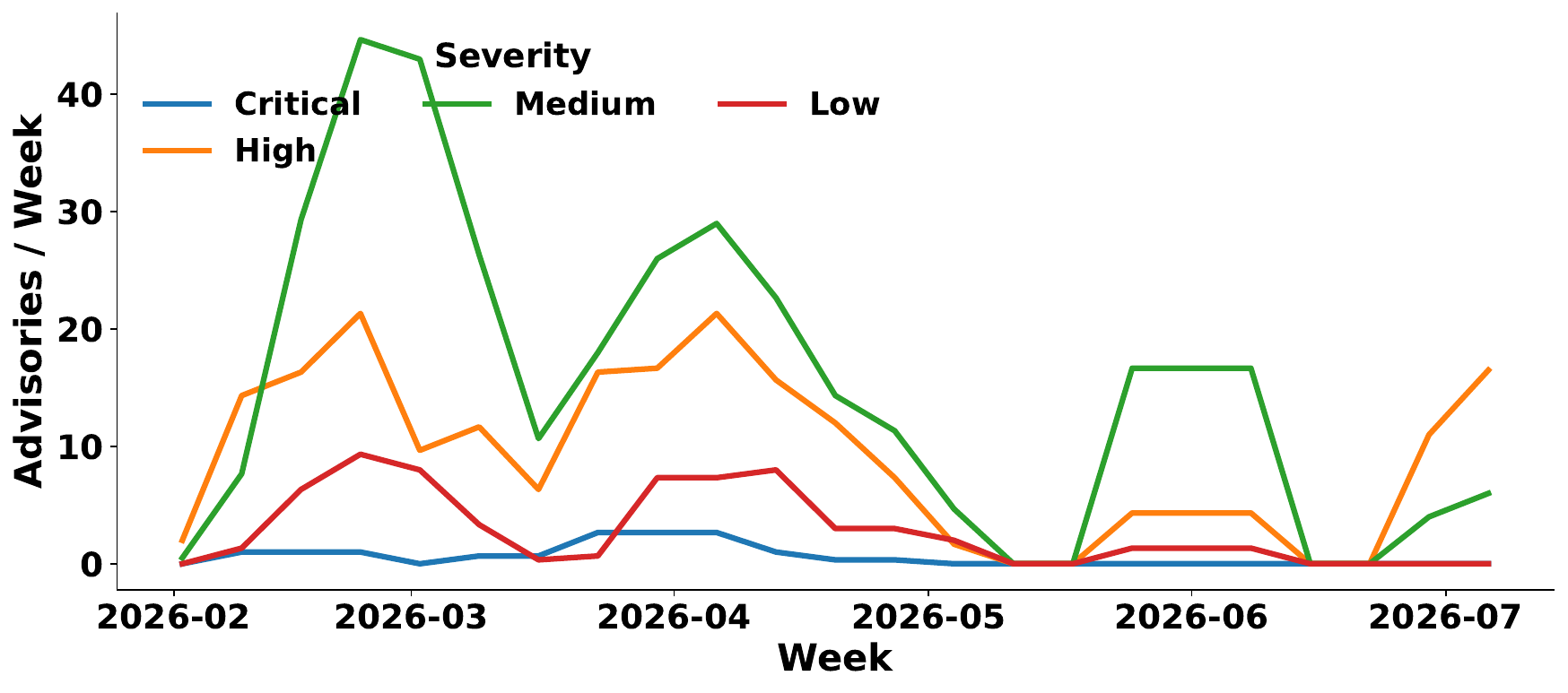}
    \caption{Weekly advisory distribution by severity.}
    \label{fig:weekly_bar}
\end{subfigure}
\hfill
\begin{subfigure}[t]{0.32\textwidth}
    \centering
    \includegraphics[width=\linewidth]{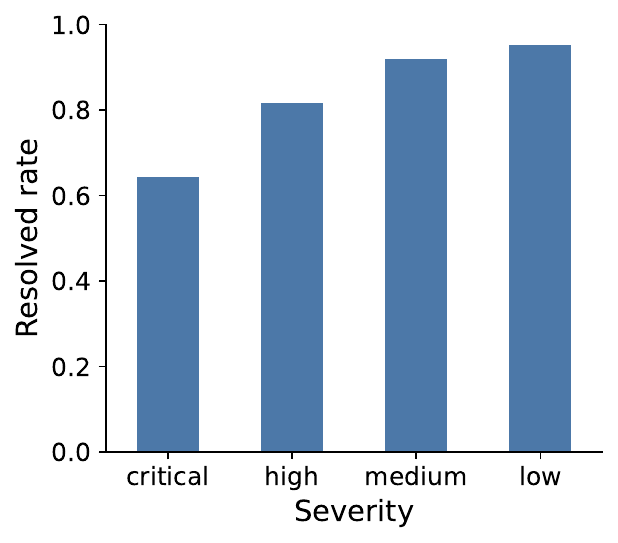}
    \caption{Resolution rate by severity.}
    \label{fig:resolve_rate}
\end{subfigure}
\hfill
\begin{subfigure}[t]{0.32\textwidth}
    \centering
    \includegraphics[width=\linewidth]{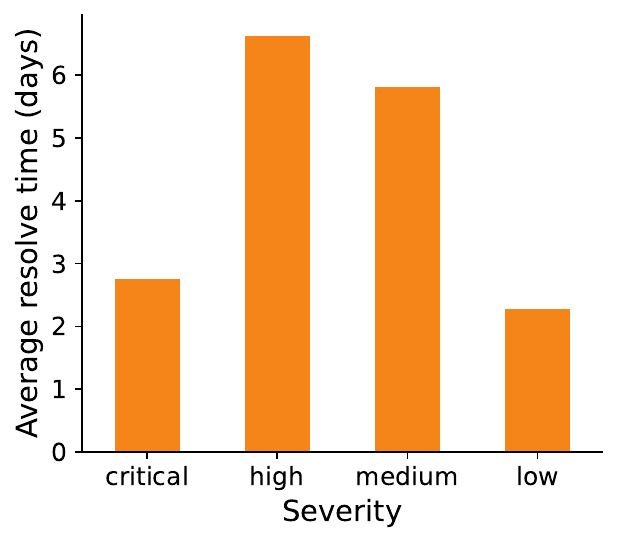}
    \caption{Average resolution time by severity.}
    \label{fig:resolve_time}
\end{subfigure}

\caption{GitHub security advisory trends for OpenClaw, including weekly severity distribution, resolution rate, and average resolution time by severity.}
\label{fig:advisory_trends}
\vspace{-0.6em}
\end{figure*}

\noindent\ul{\textbf{OpenClaw security is evolving and structurally interconnected.}}
Security issues in OpenClaw are neither static nor isolated. Advisory and CVE evidence in the supplementary material indicates continuing reports and uneven resolution patterns across severity levels. Figure~\ref{fig:advisory_trends} summarizes these trends, showing weekly advisory activity, severity-wise resolution rate, and average resolution time. These trends suggest that prioritization cannot rely on severity alone; it must also consider how weaknesses interact across access control, execution, filesystem access, network communication, and monitoring. The combined CWE, AVID, and agent-specific mechanism analyses show that OpenClaw operates within an evolving threat landscape where small configuration or validation gaps can propagate through the agent pipeline. Continuous monitoring, transparent triage, and system-level mitigation are therefore necessary for safer deployment.

\section{Threats to Validity}
\label{sec:threats_validity}

\noindent\textbf{Construct validity.}
Our analysis relies on public Reddit reports and GitHub issues, which may lack complete configuration details, exploit evidence, or root-cause information. We therefore treat them as reports of security-relevant behavior, concerns, misconfigurations, and implementation weaknesses rather than confirmed exploitable vulnerabilities. We mitigate this risk through evidence-grounded CWE mapping, conservative ambiguity handling, cross-source comparison, and advisory/CVE context where available.

\noindent\textbf{Internal validity.}
LLM-assisted filtering may miss relevant Reddit posts. To reduce this risk, we used a frozen recall-oriented prompt, two independent open-weight models, manual validation of all LLM-positive candidates, and filter audits. The audited samples showed no false negatives, and a separate audit of 200 posts excluded by both models found no missed relevant posts. Because these audits are sample-based, our counts should be interpreted as lower-bound estimates rather than exhaustive coverage.

\noindent\textbf{Coding validity.}
CWE assignment and mitigation coding involve judgment, especially for low-detail Reddit reports and practitioner comments. We mitigate this through source-aware mapping rules, conservative ambiguity handling, two-annotator review, third-author resolution, and documented coding criteria. Across the three validation tasks, exact agreement was 96.6\%, with Cohen's $\kappa$ ranging from 0.875 to 0.920. AVID is used only as a secondary interpretive layer, not as an official vulnerability assignment or replacement for CWE.

\noindent\textbf{External and temporal validity.}
Our findings are limited to OpenClaw and to public Reddit and GitHub sources, which may overrepresent visible failures and exclude enterprise deployments, private disclosures, or silent failures. The study is also time-bound because OpenClaw and its ecosystem evolve rapidly. We mitigate these risks by combining two structurally different public sources, interpreting results as a snapshot of reported concerns rather than statistically representative prevalence, and releasing anonymized replication materials.

\section{Concluding Remarks}
\label{sec:conclusion}

We presented an empirical study of reported security concerns in OpenClaw using Reddit discussions and security-tagged GitHub issues. Our analysis covers 162 security-relevant Reddit posts with 1,787 associated comments and 180 validated GitHub issues. Through evidence-grounded CWE mapping, we identified 521 CWE instances across 59 distinct weakness types, with recurring concentrations in access control, sensitive-data exposure, injection, and unsafe execution. These results show that OpenClaw security concerns are shaped by recurring trust-boundary failures rather than isolated implementation bugs. Beyond cataloging conventional weaknesses, we synthesize agent-specific failure mechanisms showing how familiar issues become amplified when model reasoning, persistent memory, tool invocation, third-party skills, credentials, and local execution interact within the same pipeline. We also analyze 543 practitioner-proposed mitigation suggestions and consolidate them into seven defense strategies, including least privilege, isolation, monitoring, supply-chain review, credential management, human approval, and network restriction. Together, these findings suggest that safer OpenClaw-like systems require defense-in-depth controls that are built into the agent pipeline, rather than left only as deployment advice.

\begin{takeaway}
\textbf{Message to OpenClaw Designers and Developers:}
Security should be enforced across the full agent execution pipeline. OpenClaw-like systems should scope permissions at action time, separate trusted instructions from untrusted data, constrain tool invocation, isolate execution contexts, and provide auditable traces for high-risk actions.
\end{takeaway}

\begin{takeaway}
\textbf{Message to Users:}
OpenClaw should be treated as a long-running software system with persistent state, external inputs, and access to sensitive resources. Users should restrict permissions, avoid public exposure of agent interfaces, isolate deployments, scope credentials carefully, and review high-risk actions before execution.
\end{takeaway}

\begin{takeaway}
\textbf{Message to Researchers:}
Future work should evaluate agentic systems end to end, rather than only testing model outputs or isolated tools. In particular, controlled reproduction of reported pathways can help determine exploitability, identify root causes, and guide concrete fixes for trust-boundary failures involving context, tools, memory, plugins, and deployment configuration.
\end{takeaway}

\section*{Ethics/AI Usage}
This study uses only public, non-interactive Reddit and GitHub data, reported in anonymized or aggregated form with sensitive details omitted; no formal IRB review was sought. Generative AI tools were used only for writing, LaTeX, visual, and coding support, not to generate experimental data, assign final labels, or produce core findings; all claims and results were author-verified.

\section*{Data Availability}
The anonymized replication package is available at: \url{https://anonymous.4open.science/r/openclaw-6E1B}. It includes filtered and annotated datasets, CWE mappings, processing scripts, filtering prompts, and coding criteria.
\bibliographystyle{ACM-Reference-Format}

\bibliography{main}

\end{document}